# Four dimensional phase unwrapping of dynamic objects in digital holography

**GILI DARDIKMAN,**[1,2] **GYANENDRA SINGH,**[1,2] **AND NATAN T. SHAKED**[1,*]

[1]*Department of Biomedical Engineering, Faculty of Engineering, Tel Aviv University, Tel Aviv 69978, Israel*
[2]*These authors contributed equally*
*\*nshaked@tau.ac.il*

**Abstract:** We present a new four-dimensional phase unwrapping approach for time-lapse quantitative phase microscopy, which allows reconstruction of optically thick objects that are optically thin in a certain temporal point and angular view. We thus use all four dimensions of the dynamic quantitative phase profile acquired, including the angular dimension and the temporal dimension, in addition to the *x-y* dimensions. We first demonstrate the capabilities of this algorithm on simulative data, enabling the quantification of the reconstruction quality relative to both the ground truth and existing unwrapping approaches. Then, we demonstrate the applicability of the proposed four-dimensional phase unwrapping algorithm by experimentally capturing a dual-angular dynamic off-axis hologram with simultaneous recording of two angular views, using multiplexing of two off-axis holograms into a single multiplexed hologram.

## References and links

1. B. Rappaz, E. Cano, T. Colomb, J. Kühn, C. Depeursinge, V. Simanis, P. J. Magistretti, and P. Marquet, "Noninvasive characterization of the fission yeast cell cycle by monitoring dry mass with digital holographic microscopy," J. Biomed. Opt. **14**(3), 034049 (2009).
2. G. Popescu, Y. Park, N. Lue, C. Best-Popescu, L. Deflores, R. R. Dasari, M. S. Feld, and K. Badizadegan, "Optical imaging of cell mass and growth dynamics," Am. J. Physiol. Cell Physiol. **295**(2), C538–C544 (2008).
3. Y. Jang, J. Jang, and Y. Park, "Dynamic spectroscopic phase microscopy for quantifying hemoglobin concentration and dynamic membrane fluctuation in red blood cells," Opt. Express **20**(9), 9673–9681 (2012).
4. Y. Bishitz, H. Gabai, P. Girshovitz, and N. T. Shaked, "Optical-mechanical signatures of cancer cells based on fluctuation profiles measured by interferometry," J. Biophotonics **7**(8), 624–630 (2014).
5. P. Girshovitz and N. T. Shaked, "Fast phase processing in off-axis holography using multiplexing with complex encoding and live-cell fluctuation map calculation in real-time," Opt. Express **23**(7), 8773–8787 (2015).
6. P. Girshovitz and N. T. Shaked, "Real-time quantitative phase reconstruction in off-axis digital holography using multiplexing," Opt. Lett. **39**(8), 2262–2265 (2014).
7. D. C. Ghihlia and M. D. Pritt, Two-Dimensional Phase Unwrapping: Theory, Algorithms, and Software (Wiley, 1998).
8. K. Itoh, "Analysis of the phase unwrapping algorithm," Appl. Opt. **21**(14), 2470 (1982).
9. J. M. Huntley and H. Saldner, "Temporal phase-unwrapping algorithm for automated interferogram analysis," Appl. Opt. **32**(17), 3047–3052 (1993).
10. M. Costantini, F. Malvarosa, F. Minati, L. Pietranera, and G. Milillo, "A three-dimensional phase unwrapping algorithm for processing of multitemporal SAR interferometric measurements," in *Proceedings of IEEE Conference on Geoscience and Remote Sensing Symposium* (IEEE, 2002), pp. 1741–1743.
11. G. Dardikman, S. Mirsky, M. Habaza, Y. Roichman, and N. T. Shaked, "Angular phase unwrapping of optically thick objects with a thin dimension," Opt. Express **25**(4), 3347–3357 (2017).
12. E. Barnhill, P. Kennedy, C. L. Johnson, M. Mada, and N. Roberts, "Real-time 4D phase unwrapping applied to magnetic resonance elastography," Magn. Reson. Med. **73**(6), 2321–2331 (2015).
13. B. J. Brewer, E. Chlebowicz-Sledziewska, and W. L. Fangman, "Cell cycle phases in the unequal mother/daughter cell cycles of Saccharomyces Cerevisiae," Mol. Cell. Biol. **4**(11), 2529–2531 (1984).
14. M. A. Herráez, D. R. Burton, M. J. Lalor, and M. A. Gdeisat, "Fast two-dimensional phase-unwrapping algorithm based on sorting by reliability following a noncontinuous path," Appl. Opt. **41**(35), 7437–7444 (2002).

## 1. Introduction

Time-lapse quantitative phase microscopy enables the study of cellular dynamics, including cell migration, swelling, lifecycle analysis and dry mass monitoring [1,2], as well as tracking the fluctuations of biological cells over time [3–5], all without using cell staining. The reconstruction process mandates the retrieval of the 2-D quantitative phase map from each of the interferograms recorded. While retrieving a phase profile from an interferogram is a well-addressed problem [5,6], the phase obtained is wrapped [7], and for optically thick objects, where the spatial gradients are steep [8], the unwrapped phase profile cannot be properly retrieved using conventional 2-D unwrapping phase algorithms, resulting in an erroneous estimation of each of the final phase maps due to $2\pi$ ambiguities.

In the field of synthetic aperture radar (SAR) interferometry, where the data usually consists of a stack of time-consecutive interferograms, it was previously suggested to utilize the time dimension in order to perform 1-D phase unwrapping along the time axis rather than 2-D phase unwrapping for each interferogram, to allow proper unwrapping even in cases of steep gradients in the phase maps [9]. This concept was later developed to 3-D phase unwrapping in a 3-D space-time domain [10].

Recently, we suggested utilizing interferometric projections acquired from consecutive angles as a third dimension for phase unwrapping, while using a simple algorithm combining 1-D phase unwrapping with 2-D phase unwrapping [11]. This method, however, can only work if two basic conditions are satisfied: sufficiently small angular increments between interferograms, and the existence of a proper boundary condition, meaning that one of the angular views is optically-thin enough to allow its phase unwrapping by conventional methods. While the former condition can usually be achieved by adjusting the experimental setup, the latter condition depends on the basic geometry of the object. If such angular view does not exist, the algorithm becomes inapplicable.

In this paper, we further generalize this approach and show that if this angular information is acquired over time, we can incorporate temporal and angular phase unwrapping in order to properly unwrap all phase maps. This allows proper phase unwrapping even for time points when the second condition of a thin angular view is not met, given that it is met at least once during the recording of the dynamic object, and that the time and angular steps between acquisitions are sufficiently small. In this new approach, we use four dimensions: the two spatial dimensions of the projection ($x$ and $y$), the illumination angle, and time. This is inherently different from previous approaches for 4-D phase unwrapping applied to resonance elastography [12], where phase is attributed to a slice ($x$, $y$ plane at certain $z$) rather than to a projection, and thus the third dimension is depth rather than viewing angle. We present two algorithms suitable for different scenarios, depending on the dimension for which the gradients are lower, and demonstrate their advantages by applying them to both simulative and experimental data and comparing them to previous approaches, showing their superiority.

## 2. Algorithm

The building block of the algorithms suggested here is a generalization of our previous combined angular-spatial phase unwrapping algorithm [11]. Practically, we can use this previous algorithm for applying phase unwrapping either with the angular or temporal dimension as a third dimension, given that two basic conditions are satisfied. These include the existence of a phase map in the sequence that can be conventionally unwrapped (thus acting as a boundary condition), and sufficiently small gradients in the selected dimension, as can be formulated by:

$$|\Delta v| \cdot \left| \frac{d\psi(x,y;v)}{dv} \right| \leq \pi, \qquad (1)$$

where $v$ is the time or angle, $\Delta v$ is the temporal or angular increment, $\psi(x,y;v)$ is the 2D phase map, and $|d\psi(x,y;v)/dv|$ is the smoothness term. For the full details and mathematical formulation, the reader is referred to our previous publication [11].

The input to the combined phase unwrapping algorithm used as a building block is $N$ wrapped phase angular or temporal projections, the $N$ coinciding initially unwrapped phase projections, and the $N$ coinciding binary masks (post registration). Then, the algorithm is defined by the following steps:

Step 1: Arrange $N$ phase maps in consecutive array, where the optically thin map (acting as a boundary condition) is unwrapped using the initial method and all the rest are wrapped: {Optically-thin *unwrapped* phase map, …, optically-thick wrapped phase map}. Then, for $k = 2,...,N$ repeat the following steps 2-4:

Step 2: Angular or temporal phase unwrapping: Apply 1-D phase unwrapping to the $(i, j)$ pixel in the $k$'th map relative to the $(i, j)$ pixel in the $(k-1)$'th map, by integrating phase differences [8].

Step 3: For each pixel in the $k$'th phase map, take the maximum value out of the initial and the angular or temporal unwrapping methods.

Step 4: Multiply by the coinciding binary mask and update the $k$'th phase map in the array.

The chaining of this algorithm with the temporal and angular dimensions enables utilization of all four dimensions for solving the phase unwrapping problem. The order of utilizing the different dimensions creates two different algorithms: the temporal/angular (T/A) algorithm and the angular/temporal (A/T) algorithm (Fig. 1).

The input to the T/A algorithm is $N\times T$ wrapped phase projections ($N$ angular views over $T$ time steps); then, the algorithm is defined by the following steps:

Step 1: Apply conventional 2-D phase unwrapping ($N\times T$ times)

Step 2: Create a binary mask and perform registration by the center of mass ($N\times T$ times)

Step 3: For each of the $N$ angular views, perform the combined unwrapping algorithm with the temporal dimension [blue arrows in Fig. 1(a)], where the initial unwrapping method is the 2D phase unwrapping from step 1.

Step 4: For each of the $T$ time steps, perform the combined unwrapping algorithm with the angular dimension [red arrows in Fig. 1(a)], where the initial unwrapping method is the temporal phase unwrapping from step 3.

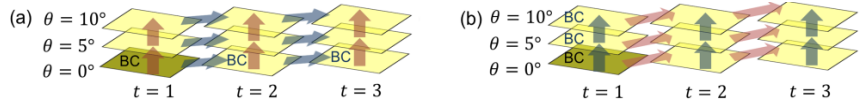

Fig. 1. (a) T/A algorithm (b) A/T algorithm. The dark rectangle in the bottom left indicates the optically thin phase map, which acts as the initial boundary condition (BC). Blue arrows are followed first, creating BCs for further unwrapping in the direction of the red arrows.

The A/T algorithm is similar to the T/A algorithm, but with 2 minor changes; in step 3, the unwrapping is performed for each of the $T$ time steps with the angular dimension [blue arrows in Fig. 1(b)]; and in step 4, the unwrapping is performed for each of the $N$ angular views with the temporal dimension [red arrows in Fig. 1(b)], where the initial unwrapping method is the angular phase unwrapping from step 3.

The order in which we place the building blocks is critical, since in each method we may or may not stumble across aliasing problems, depending on the sampling rate and sample smoothness in that dimension. For example, the A/T algorithm would be more suitable than the T/A algorithm in a scenario where for the first time step there is an optically thin phase map only from one angular view and the angular gradient is sufficiently low [11], while for all other time steps the angular gradient is high, but the temporal gradient is low for all angular views. Note that step 3 in the combined unwrapping algorithm assumes that the phase unwrapping is applied from the optically thin angle or time to the optically thick one. If the phase profile sequence is not monotonic, it should be broken down into monotonic subsequences that follow this rule.

## 3. Results

To demonstrate the improved performance of the algorithms for optically thick objects, we first conducted a simulation imitating a time-lapse off-axis interferometric tomographic system, where light was assumed to travel in straight lines, the angular increment was 5° over an angular range of 25°, and the temporal increment was 1 minute over a time frame of 19 minutes. The input to the simulation was the refractive index (RI) distribution of a 4-D test target imitating a yeast cell in suspension during reproduction by budding, such that from the 0° viewing angle (defined here as 'the optically-thin angle') for the first time step (defined here as 'the optically-thin time step'), the $x$-$y$ phase gradient is low enough to allow conventional 2-D phase unwrapping, while for other viewing angles and times the $x$-$y$ gradient is too steep to allow the retrieval of the actual phase profile. The temporal increments used in this simulation were based on a typical time frame for yeast cell reproduction by budding [13].

In Fig. 2, we show the actual phase profiles, calculated directly from the input RI distribution, in comparison to the results obtained from the wrapped phase profiles using either reliability-based 2-D phase unwrapping performed separately on each phase profile [14], angular phase unwrapping performed on each time stack separately [11], temporal phase unwrapping performed for each angular view separately, the T/A algorithm [Fig. 1(a)], the A/T algorithm [Fig. 1(b)], or three different types of previously suggested 4-D phase unwrapping algorithms [12]. The latter include a Laplacian-based estimate (LBE) algorithm, a 4-D version of the reliability-based algorithm, also called the region-growing algorithm (RG), and the dilate-erode-propagate (DE) algorithm, all implemented in the PhaseTools application created by Barnhill et al. [12]. For each method, we show chosen phase profiles taken from the 0° viewing angle (optically-thin angle) and from the 25° viewing angle, at $t = 1$ min (optically-thin time) as well as at $t = 12$ min.

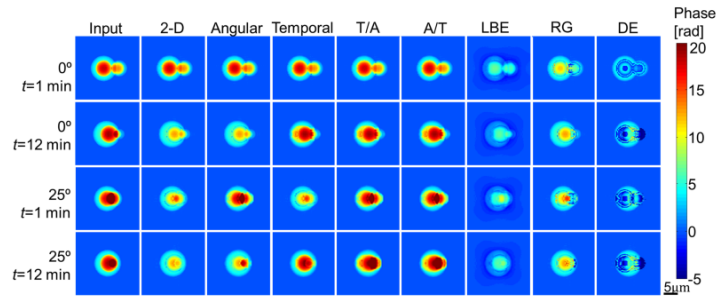

Fig. 2. Phase reconstruction using different phase unwrapping algorithms, applied on simulative data. First column to the left: actual phase profile (Visualization 1). Second column: 2-D reliability-based phase unwrapping performed separately on each phase profile [14] (Visualization 2). Third column: angular phase unwrapping performed for each time stack separately [11] (Visualization 3). Fourth column: temporal phase unwrapping performed for each angular view separately (Visualization 4). Fifth column: T/A algorithm (Visualization 5). Sixth column: A/T algorithm (Visualization 6). Seventh column: LBE Algorithm [12] (Visualization 7). Eighth column: RG Algorithm [12] (Visualization 8). Ninth column: DE Algorithm [12] (Visualization 9). First row: optically-thin time and angle (boundary condition for the T/A and A/T algorithms). Note that in the visualizations, the time steps are shown from the end to the beginning, to show the budding process, where the last time step (shown here as the first) acts as the optically-thin time step.

The mean squared error (MSE) was calculated for each unwrapping method relative to the actual phase profile over 6 viewing angles in 19 time steps, and was 2.55 rad$^2$ for the 2-D algorithm, 1.70 rad$^2$ for the angular algorithm, 0.95 rad$^2$ for the temporal algorithm, 0.79 rad² for the T/A algorithm, 0.82 rad² for the A/T algorithm, 8.68 rad² for the LBE algorithm, 6.15 rad² for the RG algorithm, and 12.79 rad² for the DE algorithm. Note that the relatively high MSE obtained for all algorithms is a result of the abnormally high gradients intentionally created in this simulation for all dimensions, in order to examine the algorithms for an extreme scenario. In any case, the A/T and T/A algorithms yielded similar results for this data set, and are both superior to all previous methods. The previous methods tested fail in different locations, each for its own reasons; the 2-D unwrapping method fails for phase maps with steep $x$-$y$ gradients, which are all phase maps other than the one taken at the optically thin time and angle; the angular algorithm fails for angular stacks without a proper boundary condition, which are all angular stacks not taken at the optically thin time $t = 1$; the temporal algorithm similarly fails for temporal stacks without a proper boundary condition, which are all temporal stacks not taken at the optically thin angle $\theta = 0°$; and all 4-D phase unwrapping algorithms fail far worse than the 2-D phase unwrapping algorithm for all phase maps. For the LBE algorithm, this can be explained by the fact that it is never perfect even in trivial cases, and is accurate only when the overall phase range is not much greater than $[-\pi, \pi)$; for the RG algorithm, rapid changes in gradient can cause its region-forming criterion to err, especially along more coarsely sampled dimensions; and the DE algorithm is optimal for data with moderate or low gradients that is disrupted by high noise, and functions as a standalone unwrapper only where at least one slice has sufficiently light wrap such that all wrapped clusters are smaller than the window used for unwrapping [12].

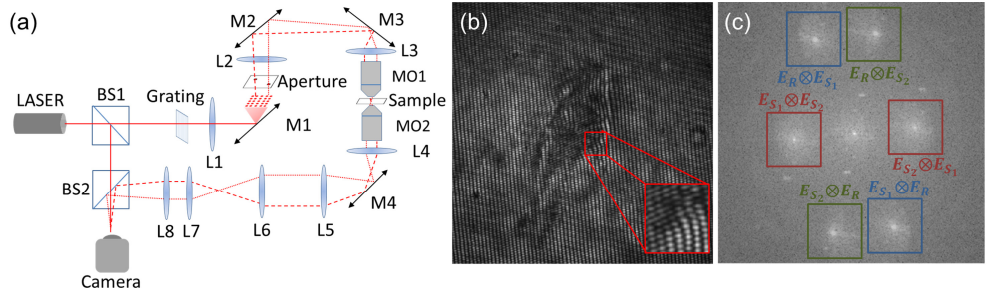

Fig. 3. Experimental demonstration: (a) Mach-Zehnder-based off-axis angular multiplexing interferometer, used for data acquisition. BS1, BS2: beam splitters; L1, L2: lenses; L3-L8: lenses, with each pair forming a 4f lens configuration, including MO1 and MO2; M1-M4: mirrors; MO1, MO2: microscope objectives. (b) Multiplexed off-axis hologram acquired with this setup at $t = 0$ sec. (c) Absolute value of the Fourier transform of the multiplexed hologram, featuring three different cross-correlation pairs in non-overlapping locations. $E_R$: Fourier transform of reference beam wavefront, $E_{s_1}$: Fourier transform of the -7° sample beam wavefront, $E_{s_2}$: Fourier transform of the +21° sample beam wavefront. $\otimes$ indicates the cross-correlation operator.

In order to allow simultaneous recording of two angular views for each time step, we built an optical setup combining a conventional off-axis Mach-Zehnder interferometer with angular multiplexing, as shown in Fig. 3(a). In this setup, a He-Ne laser (HNL210-L, Thorlabs, 632.8 nm) is split using a 50:50 beam splitter (BS1). One beam passes through a 2-D grating and lens L1 (f = 250mm), and is then reflected by mirror M1, yielding a projection of the grating pattern on an aperture. Out of the grating diffraction orders, only two beams are passed by the aperture, illuminating the sample from two different angles, thus acting as two illumination perspective sample beams. The two beams intersect using lens L2 (f = 250 mm), and are then projected onto lens L3 (f = 60 mm) using mirrors M2 and M3. Lens L3 is positioned close to the back focal plane of the first microscope objective (MO1, Newport 60×, numerical aperture 0.85). The collimated beams then pass through MO1 and intersect at the sample plane, each illuminating it from a different angle. After passing through the sample, the beams are collected by MO2 (Newport 60×, numerical aperture 0.65) and collimated by lens L4. The 60× magnified image plane is then projected onto the monochromatic CMOS digital camera (DCC1545, Thorlabs, with 1024 × 1280 square pixels, 5.2 µm each) using lenses L5 (f = 180 mm), L6 (f = 200 mm), L7 (f = 25.4 mm), and L8 (f = 125 mm), where each pair of these lenses is positioned in a 4f lens configuration. After being combined by beam splitter BS2, the two sample beams interfere with the reference beam on the camera plane in different off-axis angles, creating a three-beam interference with three different fringe orientations [see Fig. 3(b)], forming three non-overlapping cross-correlation pairs in the spatial frequency domain, out of which two pairs are relevant [see Fig. 3(c)].

We used this setup to image micro-organisms (euglena gracilis) in a viscous medium between two coverslips (10% solution polyvinylpyrrolidone (PVP)). We recorded a video of a dynamic multiplexed off-axis hologram at a framerate of 6 fps. The wrapped phase profiles were reconstructed using the typical Fourier transform filtering reconstruction algorithm for off-axis holograms [6], prior to applying the various phase unwrapping algorithms, with the results given in Fig. 4. As can be seen from Fig. 4, even though the angular increment was much more significant than the temporal one, the A/T and T/A algorithms yield similar results, and are both superior to previous methods.

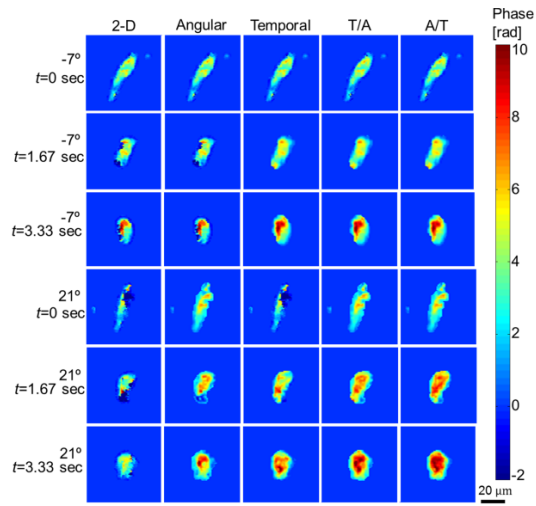

Fig. 4. Phase reconstruction using different phase unwrapping algorithms, applied on experimental data of a dynamic micro-organism. First column to the left: 2-D reliability-based phase unwrapping performed on each phase profile separately [14] (Visualization 10). Second column: angular phase unwrapping performed on each time stack separately [11] (Visualization 11). Third column: temporal phase unwrapping performed on each angular view separately (Visualization 12). Fourth column: T/A algorithm [presented in Fig. 1(a)] (Visualization 13). Fifth column: A/T algorithm [presented in Fig. 1(b)] (Visualization 14). First row: optically-thin time and angle phase profiles (boundary condition for the T/A and A/T algorithms).

## 4. Conclusions

The new approaches for temporal/angular and angular/temporal phase unwrapping presented in this paper enable phase unwrapping of optically thick objects captured over time from multiple angles, even if for all time and angular steps other than one there is no optically-thin phase map to act as a boundary condition. In this paper, we demonstrated the acquisition of two angular views per each time step. Nevertheless, the suggested 4-D algorithm is also useful for time-lapse tomographic phase microscopy, where multiple angular views are acquired over time, provided that the angular scanning is faster than the sample dynamics. In such case, the use of the angular dimension may allow solving phase ambiguities for angular views that could not be solved using the spatial and temporal dimensions alone due to a lack of boundary condition (A/T algorithm), and vice versa (T/A algorithm).

**Funding**

Horizon 2020 European Research Council (ERC) Grant No. 678316; Tel Aviv Center for Light Matter Interaction.